\documentclass[runningheads]{llncs}
\usepackage{graphicx}
\usepackage{amsfonts}
\usepackage{amsmath}
\usepackage{bbm}
\usepackage{color}
\usepackage{booktabs}
\usepackage{multirow}
\usepackage{xspace}
\usepackage{enumitem}
\usepackage{xcolor}
\usepackage{caption}
\usepackage{subcaption}
\usepackage[export]{adjustbox}
\usepackage[normalem]{ulem}
\usepackage[misc]{ifsym}
\useunder{\uline}{\ul}{}

\def\modelName{LAGCL} 

\newcommand{\eg}{\emph{e.g.,}\xspace}

\newcommand{\ie}{\emph{i.e.,}\xspace}
\newcommand{\ignore}[1]{}

\begin{document}
%
\title{Long-tail Augmented Graph Contrastive Learning for Recommendation}
%
%
\author{
    Qian Zhao \and
    Zhengwei Wu \and
    Zhiqiang Zhang \and
    Jun Zhou{\textsuperscript{\Letter}}
}
\tocauthor{
    Qian~Zhao,
    Zhengwei~Wu,
    Zhiqiang~Zhang and
    Jun~Zhou{\textsuperscript{\Letter}}
}
\toctitle{Long-tail Augmented Graph Contrastive Learning for Recommendation}

%
%
\institute{Ant Group, Hangzhou, China \\
\email{\{zq317110, zejun.wzw, lingyao.zzq, jun.zhoujun\}@antgroup.com}}
\maketitle              
\begin{abstract}

Graph Convolutional Networks (GCNs) has demonstrated promising results for recommender systems, as they can effectively leverage high-order relationship. However, these methods usually encounter data sparsity issue in real-world scenarios. To address this issue, GCN-based recommendation methods employ contrastive learning to introduce self-supervised signals. Despite their effectiveness, 
these methods lack consideration of the significant degree disparity between head and tail nodes. 
This can lead to non-uniform representation distribution, which is a crucial factor for the performance of contrastive learning methods.
To tackle the above issue, we propose a novel \textbf{L}ong-tail \textbf{A}ugmented \textbf{G}raph \textbf{C}ontrastive \textbf{L}earning ({\modelName}) method for recommendation. Specifically, we introduce a learnable long-tail augmentation approach to enhance tail nodes by supplementing predicted neighbor information, and generate contrastive views based on the resulting augmented graph. To make the data augmentation schema learnable, we design an auto drop module to generate pseudo-tail nodes from head nodes and a knowledge transfer module to reconstruct the head nodes from pseudo-tail nodes. Additionally, we employ generative adversarial networks to ensure that the distribution of the generated tail/head nodes matches that of the original tail/head nodes.
Extensive experiments conducted on three benchmark datasets demonstrate the significant improvement in performance of our model over the state-of-the-arts. Further analyses demonstrate the uniformity of learned representations and the superiority of {\modelName} on long-tail performance.

\end{abstract}
\section{Introduction}
Recommender systems are a critical component of numerous online services, ranging from e-commerce to online advertising. As a classic approach, collaborative filtering (CF)~\cite{he2017neural,sarwar2001item} plays a vital role in personalized preference prediction by representing user and item embeddings from observed user-item interactions such as clicks and conversions. Recently, enhanced by the powerful Graph Convolutional Networks (GCNs)~\cite{kipf2016semi}, GCN-based recommendation methods~\cite{he2020lightgcn,wang2019neural} have demonstrated significant potential in improving recommendation accuracy. GCN-based recommendation methods represent interaction data as graphs, such as the user-item interaction graph, and iteratively propagate neighborhood information to learn effective node representations. Compared to traditional CF methods, GCN-based recommendation methods are better equipped to capture higher-order collaborative signals, leading to improved user and item embedding learning.

Despite the effectiveness, GCN-based recommendation methods still face data sparsity issue in real-word scenarios. Most existing models follow the supervised learning paradigm~\cite{berg2017graph,he2020lightgcn,wang2019neural,ying2018graph}, where the supervision signal is derived from the observed user-item interactions. However, the observed interactions are considerably sparse in contrast to the entire interaction space~\cite{wu2021self-2}. As a result, they may not be sufficient to acquire effective representations.
Although several recent studies have attempted to alleviate the data sparsity of interaction data through contrastive learning~\cite{wu2021self-2,yu2022graph}, they generally rely on pre-defined data augmentation strategies, such as uniformly dropping edges or shuffling features. These methods lack consideration of the significant disparity in the graph structure between head and tail nodes and lack the ability to construct adaptive data augmentation tailored for various recommendation datasets. 
This can lead to non-uniform representation distribution, which is a crucial factor for the performance of contrastive learning methods~\cite{wang2020understanding,yu2022graph}.


\begin{figure}[h]
    \centering
    \begin{subfigure}{.26\textwidth}
            \centering
    	\includegraphics[width=0.9\textwidth, raise=.05\height]{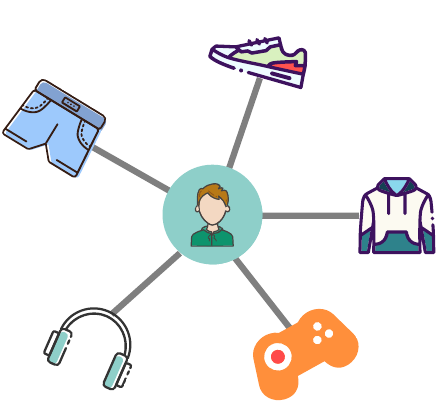}
    	\caption{}
            \label{fig:intro_demo_a}
    \end{subfigure}
    \hspace{0.05\textwidth}
    \begin{subfigure}{.26\textwidth}
            \centering
    	\includegraphics[width=0.6\textwidth, raise=.05\height]{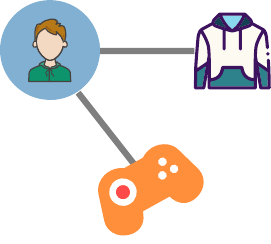}
    	\caption{}
            \label{fig:intro_demo_b}
    \end{subfigure}
    \hspace{0.05\textwidth}
    \begin{subfigure}{.26\textwidth}
            \centering
    	\includegraphics[width=\textwidth]{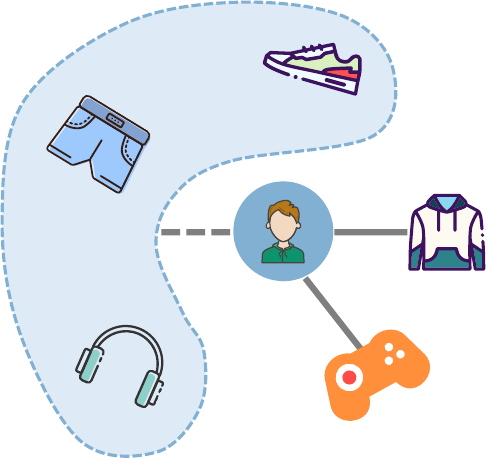}
    	\caption{}
            \label{fig:intro_demo_c}
    \end{subfigure}
    \caption{An illustrated example depicting (a) a head node, (b) a tail node, and (c) a tail node augmented with predicted neighbor information.}
    \label{fig:intro_demo}
\end{figure}

In the light of the above limitations and challenges, we propose a novel \textbf{L}ong-tail \textbf{A}ugmented \textbf{G}raph \textbf{C}ontrastive \textbf{L}earning ({\modelName}) method for recommendation. To illustrate, consider the head user in Fig.~\ref{fig:intro_demo}(a) and the tail user in Fig.~\ref{fig:intro_demo}(b) who share similar preference. Our approach aims to extract informative transition patterns from head users and adapt to tail users effectively, as shown in Fig.~\ref{fig:intro_demo}(c). 
Specifically, we first design an auto drop module to convert head nodes into pseudo-tail nodes that mimic the patterns of real-tail nodes.
Next, we leverage the dropped neighbor information of pseudo-tail nodes to learn the knowledge transfer module, which then augments the tail nodes as pseudo-head nodes by adding predicted neighbor information.
These modules are updated using the adversarial training mechanism to ensure the distribution match between pseudo-head/tail nodes and real-head/tail nodes. 
Finally, we use an effective and efficient approach to generate contrastive views by injecting random noise into the graph embedding of both head nodes and augmented tail nodes, capitalizing on their uniformity to yield better contrastive performance.

The main contributions of this paper are summarized as follows:
\begin{itemize}
    \item We propose a novel graph contrastive learning method, which encourages the model to learn the knowledge between head and tail nodes and generate uniform representation distribution for improving the GCN-based recommendation.
    \item We designed a learnable data augmentation scheme that can adaptively enhance tail nodes representation and easily generalize to different GCN-based recommendation scenarios.
    \item Extensive experiments are conducted on three public datasets, demonstrating our approach is consistently better than a number of competitive baselines, including GCN-based and graph contrastive learning-based recommendation methods.
\end{itemize}


\section{Preliminaries}
\subsection{GCN-based Recommendation}
Given the user set $\mathcal{U}=\{u\}$ and the item set $\mathcal{I}=\{i\}$, the observed user-item interaction data is denoted as $\mathbf{R}\in{\mathbb{R}}^{|\mathcal{U}|\times|\mathcal{I}|}$, where each entry $r_{u,i}=1$ if there exists an interaction between user $u$ and item $i$, otherwise $r_{u,i}=0$. The number of nodes is $n=|\mathcal{U}|+|\mathcal{I}|$.
GCN-based recommendation methods formulate the available data as a user-item bipartite graph $\mathcal{G}=(\mathcal{V}, \mathbf{A})$, where $\mathcal{V}=\mathcal{U} \cup \mathcal{I}$ and $\mathbf{A} \in\mathbb{R}^{n\times n}$ is the adjacent matrix defined as
\begin{equation}
  \mathbf{A}=\begin{bmatrix}
    \mathbf{0}^{|\mathcal{U}| \times |\mathcal{U}|} & \mathbf{R}                                      \\
    \mathbf{R}^T                                    & \mathbf{0}^{|\mathcal{I}| \times |\mathcal{I}|}
  \end{bmatrix}.
\end{equation}
With a slight abuse of notation, we use $|\mathbf{A}_i|$ to refer to $\sum_{j\in\mathcal{N}_i}\mathbf{A}_{ij}$, where $\mathcal{N}_i$ denotes the neighbor set of node $i$. GCN-based recommendation methods utilize graph structure information to aggregate and produce the embedding of users and items on bipartite graph $\mathcal{G}$ through Eq. (\ref{eq:lightgcn_agg}).
\begin{equation}
  \mathbf{H}^{(l)} = \mathbf{D}^{-\frac{1}{2}}\mathbf{A}\mathbf{D}^{-\frac{1}{2}}\mathbf{H}^{(l-1)},
  \label{eq:lightgcn_agg}
\end{equation}
where $\mathbf{D}\in\mathbb{R}^{n\times n}$ is the diagonal degree matrix of $\mathcal{G}$,
in which each entry $\mathbf{D}_{ii}$ denotes the number of non-zeros in the $i$-th row of the matrix $\mathbf{A}$. $\mathbf{H}^{(l)}\in\mathbb{R}^{n\times d}$ denotes the $d$-dimensional node embedding matrix after $l$-th graph convolution layer, and $\mathbf{H}^{(0)}$ is the initial node embedding matrix that need to be learned.
Finally, we combine all the $L$ layers output node embeddings, \ie $\mathbf{H}=f_{\rm readout}([\mathbf{H}^{(0)}; \mathbf{H}^{(1)}; \cdots ;\mathbf{H}^{(L)}])$, to generate preference scores between users and items for recommendation, while $f_{\rm readout}(\cdot)$ is the mean pooling operation here.

\subsection{Long-Tail Distribution in the Graph}
Graph convolutional networks heavily rely on rich structural information to achieve high performance.
However, for nodes with low degrees, their number of neighbors is typically very small, leading to unsatisfactory performance for these nodes.
In this paper, in order to investigate the long-tail distribution in the graph, we partition nodes into head nodes $\mathcal{V}_{head}$ and tail nodes $\mathcal{V}_{tail}$ based on their degree with a predetermined threshold value $k$, i.e., $\mathcal{V}_{head}=\{i: \mathbf{D}_{ii} > k\}$ and $\mathcal{V}_{tail}=\{i: \mathbf{D}_{ii} \le k\}$. Specifically, we have $\mathcal{V}_{head} \cap \mathcal{V}_{tail} = \emptyset$ and $\mathcal{V}_{head} \cup \mathcal{V}_{tail} = \mathcal{V}$.

\section{Methodology}
\label{section:methodology}
In this section, we introduce the proposed long-tail augmented graph contrastive learning (LAGCL) model for recommendation. First, we briefly exhibit the overall architecture of LAGCL. Then, we elaborate the detail implementation of LAGCL.


\begin{figure}
  \centering
  \includegraphics[width=1.0\textwidth]{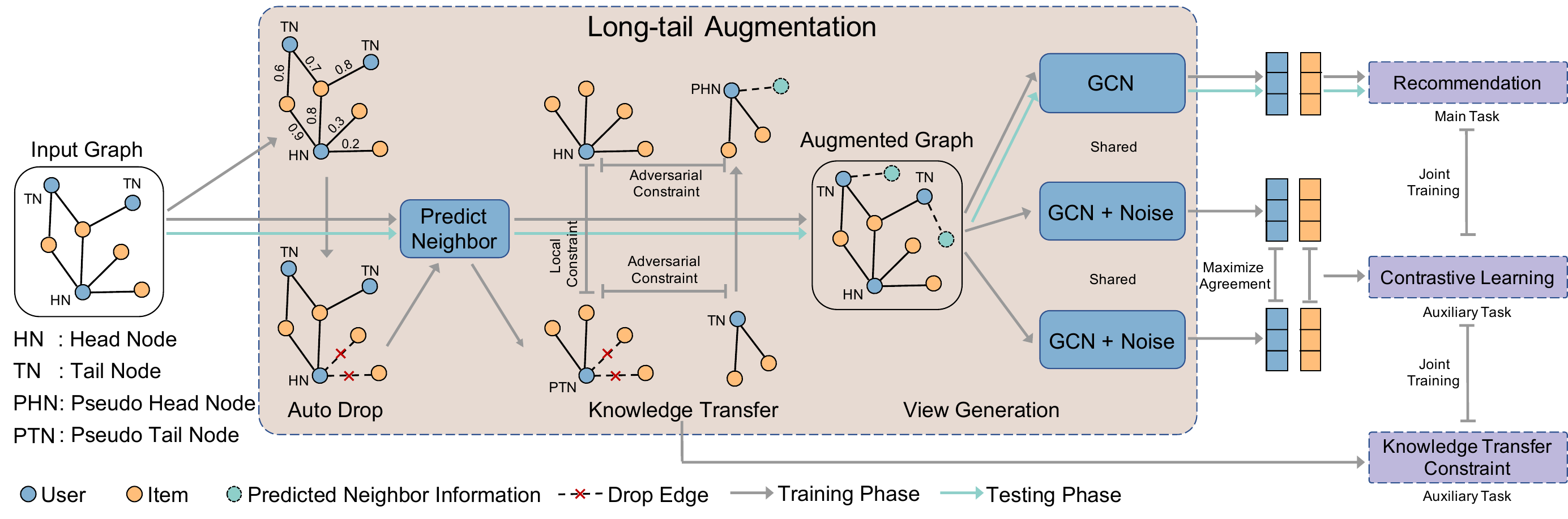}
  \caption{Overall framework of our proposed long-tail augmented graph contrastive learning method for recommendation. To enhance readability, only user-side augmentation is shown. The core of the model is long-tail augmentation, which extracts informative transition from head nodes and augments tail nodes using the knowledge transfer module. The augmented graph is then used to generate the view pairs for contrastive learning. 
  Finally, we train recommendation task, contrastive learning task and knowledge transfer constraints jointly using the multi-task training strategy.
  }
  \label{fig:pipeline}
\end{figure}

\subsection{Overview of {\modelName}}
Our framework follows the general contrastive learning paradigm, which aims to achieve maximum agreement between representations of different views. Fig.~\ref{fig:pipeline} illustrates the overall framework. Firstly, we enhance tail node by a knowledge transfer module and then generate two graph views on the augmented graph. Next, we employ a contrastive loss function to conduct the contrastive learning task, which encourages representations of the same nodes in the two different views to be similar, while representations of different nodes in those views to be distinct. Finally, we adopt a multi-task training strategy to improve recommendation performance. Specifically, the main task is the recommendation task, while contrastive learning task and knowledge transfer constraints serve as auxiliary tasks.

\subsection{Long-tail Augmentation}

Assuming that each node has a ground truth neighbor set, the difference between head and tail nodes lies in the former having a fully observed neighbor set while the latter has only partially observed neighbor information. This incomplete information adversely affects the performance of the model. Therefore, the technique of long-tail augmentation aims to supplement the tail nodes with missing neighbor information, with the aim of enforcing representation uniformity between head and tail nodes. However, conducting long-tail augmentation is a challenging task due to unavailability of ground truth labels for the missing neighbors of tail nodes. The intuition behind this approach is that head nodes have fully observed neighbor information. If we could generate pseudo-tail nodes by dropping the neighbor information of head nodes to mimic tail nodes, we could leverage the dropped neighbor information of pseudo-tail nodes to learn the long-tail augmentation strategy. Our long-tail augmentation boils down to three key modules: auto drop module, knowledge transfer module and generative adversarial learning module.

\subsubsection{Auto Drop Module.}
\label{section:auto-drop-module}


An intuitive approach is to randomly sample neighbors from head nodes to generate pseudo-tail nodes. However, this method lacks an effective supervisory signal to guide the sampling process, leading to a distribution deviation between the enhanced data samples and the actual tail node samples. To improve the data augmentation process, we propose an auto drop module equipped with a trainable dropout strategy to minimize the distribution deviation between the enhanced data samples and the actual tail node samples.
Specifically, the dropout strategy consists of two main steps: sorting by edge weight and selecting top-$K$ neighbors based on their importance weight, which is calculated between node pairs as:
\begin{equation}
  \label{eq:aw_build}
  \mathbf{S}=(\mathbf{H}^{(0)}\mathbf{W}_s {\mathbf{H}^{(0)}}^T)\odot \mathbf{A},
\end{equation}
where $\mathbf{S} \in \mathbb{R}^{n\times n}$, $\mathbf{W}_s\in\mathbb{R}^{d\times d}$ is trainable parameters and $\odot$ denotes the element-wise product.
Then, to mimic real-tail nodes, we randomly choose the neighbor size $k_i$ of head nodes uniformly from the range of tail node neighbor sizes $[1,k]$:
\begin{equation}
  \label{eq:ki_build}
  k_{i}=
  \begin{cases}
    \text{Uniform}[1,k], & \mathbf{D}_{ii} > k,   \\
    \mathbf{D}_{ii},            & \mathbf{D}_{ii} \le k.
  \end{cases}
\end{equation}
Finally, inspired by the top-rank selection method in~\cite{gao2019graph,lee2019self}, the new adjacency matrix $\mathbf{\hat{A}}$ is constructed by selecting the top-$K$ neighbors based on their importance weight, which will be used in the graph aggregation of tail nodes. The new adjacency matrix $\mathbf{\hat{A}}$ is defined as:
\begin{equation}
  \label{eq:a_tail_build}
  \hat{\mathbf{A}}_{ij}=
  \begin{cases}
    \frac{\exp(\delta \mathbf{S}_{ij})}{1 + \exp (\delta \mathbf{S}_{ij})}, & \mathbf{S}_{ij}\in \text{top-}k_i(\mathbf{S}_i), \\
    0, & \mathbf{S}_{ij}\notin \text{top-}k_i(\mathbf{S}_i),
  \end{cases}
\end{equation}
where $\delta$ is a hyperparameter that controls the smoothness of the adjacency matrix $\mathbf{\hat{A}}$, and the corresponding neighborhood of node $i$ is denoted as $\hat{\mathcal{N}}_i$. 

\subsubsection{Knowledge Transfer Module.}
\label{section:knowledge-transfer-module}

After constructing the auto drop module, we can design a knowledge transfer module that leverages the dropped neighbor information of pseudo-tail nodes for predicting the missing neighbor information of real-tail nodes.
Specifically, we use a multi-layer perceptron (MLP) function, denoted by $f_t(\mathbf{h}_i^{(l)}, \mathbf{h}_{\mathcal{N}_i}^{(l)};\theta_t^{(l)})=\mathbf{m}_i^{(l)}$, to predict the missing neighbor information based on the center node features and observed neighbor information. Here, $\mathbf{h}_i^{(l)}$ represents the $l$-layer graph embedding of node $i$, $\mathbf{h}_{\mathcal{N}_i}^{(l)}$ is the mean pooling representation of observable neighbors.
Then, the predicted information is added to the neighbor aggregation process of real-tail node.

We can calculate the node embedding of head node $i$ in the sparse graph $\hat{\mathbf{A}}$ with predicted neighbor information by the knowledge transfer function:
\begin{equation}
  \hat{\mathbf{h}}_i^{(l)} = \sum_{j\in\hat{\mathcal{N}}_i}\frac{1}{\sqrt{|\hat{\mathbf{A}}_i|}\sqrt{|\hat{\mathbf{A}}_j|}}\hat{\mathbf{h}}_j^{(l-1)} + f_t(\hat{\mathbf{h}}_i^{(l-1)}, \hat{\mathbf{h}}_{\hat{\mathcal{N}}_i}^{(l-1)}, \theta_t^{(l-1)}),
\end{equation}
where $\hat{\mathbf{h}}^{(0)}=\mathbf{h}^{(0)}$. The embedding of head node $i$ in the original graph $\mathbf{A}$ is
\begin{equation}
  \mathbf{h}_i^{(l)} = \sum_{j\in\mathcal{N}_i}\frac{1}{\sqrt{|\mathbf{A}_i|}\sqrt{|\mathbf{A}_j|}}\mathbf{h}_j^{(l-1)}.
\end{equation}
In order to train the knowledge transfer function, we define the translation loss is defined as follows:
\begin{equation}
  \label{eq:knowledge_transfer_loss}
  \mathcal{L}_{trans} = \sum_{i\in\mathcal{V}_{head}}\sum_{l=1}^L\|\mathbf{h}_i^{(l)} - \hat{\mathbf{h}}_i^{(l)}\|_2^2.
\end{equation}

\subsubsection{Generative Adversarial Learning Module.}
\label{section:long-tail-aware-contrastive-learning-module}
To learn effective long-tail augmentation, the representation distribution of real-tail nodes should match the representation distribution of pseudo-tail nodes generated by the auto drop module, which is calculated as follows:
\begin{equation}
  \tilde{\mathbf{h}}_i^{(l)} = \sum_{j\in\hat{\mathcal{N}}_i}\frac{1}{\sqrt{|\hat{\mathbf{A}}_i|}\sqrt{|\hat{\mathbf{A}}_j|}}\tilde{\mathbf{h}}_j^{(l-1)},
\end{equation}
where $\tilde{\mathbf{h}}^{(0)}=\mathbf{h}^{(0)}$. Additionally, the distribution of pseudo-head nodes augmented by the knowledge transfer module should match that of real-head nodes. 
To achieve this, we use Generative Adversarial Networks~\cite{NIPS2014_5ca3e9b1}. The discriminator distinguishes pseudo-head/tail nodes from real-head/tail nodes based on the node representations, while the generator aims to provide information that is consistently classified as real nodes by the discriminator. 
Here we regard the output layer of {\modelName} as the generator, which contests with the discriminator in the learning process. In particular, we use the following loss for the tail nodes adversarial constraint:
\begin{eqnarray}
  \label{eq:tail_disc_loss}
  \mathcal{L}_{tail-disc}=\sum_{i\in\mathcal{V}}&\mathbbm{1}(i\notin\mathcal{V}_{tail})\textsc{CrossEnt}(\mathbf{0}, f_d(\tilde{\mathbf{h}}_i;\theta_d))\nonumber \\
  &+ \mathbbm{1}(i\in\mathcal{V}_{tail})\textsc{CrossEnt}(\mathbf{1}, f_d(\mathbf{h}_i;\theta_d)),
\end{eqnarray}
and use the following loss for the head nodes adversarial constraint:
\begin{eqnarray}
  \label{eq:head_disc_loss}
  \mathcal{L}_{head-disc}=\sum_{i\in\mathcal{V}}&\textsc{CrossEnt}(\mathbf{0}, f_d(\hat{\mathbf{h}}_i;\theta_d))
  \nonumber \\
  &+ \mathbbm{1}(i\in\mathcal{V}_{head})\textsc{CrossEnt}(\mathbf{1}, f_d(\mathbf{h}_i;\theta_d)),
\end{eqnarray}
where $\textsc{CrossEnt}(\cdot)$ is the cross entropy function, $\mathbbm{1}(\cdot)$ is the indicator function, $f_d(\cdot;\theta_d)$ is the discriminator function parameterized by $\theta_d$, which calculates the probability of a node being a head node, as 
\begin{equation}
  \label{eq:disc_loss}
  f_d(\mathbf{h}_i;\theta_d)=\sigma\Big(\mathbf{w}_d^\top\textsc{LeakyReLU}(\mathbf{W}_d\mathbf{h}_i+\mathbf{b}_d)\Big),
\end{equation}
where $\textsc{LeakyReLU}(\cdot)$ is used as the activation function, $\sigma(\cdot)$ is the sigmoid function, $\theta_d=\{\mathbf{W}_d \in \mathbb{R}^{d \times d},\mathbf{b}_d \in \mathbb{R}^{d \times 1},\mathbf{w}_d \in \mathbb{R}^{d \times 1} \}$ contains the learnable parameters of the discriminator $f_d$.



\subsection{Contrastive Learning}
\label{section:multiview_contrastive_learning}
\subsubsection{View Generation.}
With the knowledge transfer function mentioned above, we can obtain the augmented tail node embedding, as follows:
\begin{equation}
  \mathbf{h}_i^{(l)} = \sum_{j\in\mathcal{N}_i}\frac{1}{\sqrt{|\mathbf{A}_i|}\sqrt{|\mathbf{A}_j|}}\mathbf{h}_j^{(l-1)}+f_t(\mathbf{h}_i^{(l-1)},\mathbf{h}_{\mathcal{N}_i}^{(l-1)};\theta_t^{(l-1)}).
\end{equation}
Then, we follow the approach of SimGCL \cite{yu2022graph} and generate different views by slightly rotating refined node embeddings in space. This method retains original information while introducing the InfoNCE loss as an additional self-supervised task to improve robustness, as follows:
\begin{equation}
  \label{eq:simgcl_noise}
  \mathbf{h}_i^{(l)'} = \mathbf{h}_i^{(l)} + \Delta_i^{(l)'}, \mathbf{h}_i^{(l)''} = \mathbf{h}_i^{(l)} + \Delta_i^{(l)''},
\end{equation}
where the noise vectors $\Delta_i'$ and $\Delta_i''$ are subject to $||\Delta||_2=\epsilon$ and $\Delta = \bar{\Delta} \odot {\rm sign}(\mathbf{h}_i^{(l)})$, $\bar{\Delta} \in \mathbb{R}^d \sim U(0, 1)$.
We can use $\epsilon$ to control the rotation angle of $\mathbf{h}_i^{(l)'}, \mathbf{h}_i^{(l)''}$ compared to $\mathbf{h}_i^{(l)}$.
Since $\mathbf{h}_i^{(l)}, \mathbf{h}_i^{(l)'}, \mathbf{h}_i^{(l)''}$ always belong to the same hyperoctant, so adding noise will not cause significant deviation.
The noise is injected into each convolution layer and we average each layer output as the final node embedding. To simply the notation, we denote $\mathbf{h}_i$ as the final node embedding after $L$ layers, $\mathbf{h}_i'$ and $\mathbf{h}_i''$ as two generated views.

\subsubsection{Contrastive Loss.}
After obtaining the generated views of each node, we utilize the contrastive loss, InfoNCE~\cite{gutmann2010noise}, to maximize the agreement of positive pairs and minimize that of negative pairs:
\begin{equation}
  \label{eq:info_nce}
  \mathcal{L}_{cl}^U = \sum_{u\in\mathcal{U}} -\log \frac{\exp(s(\mathbf{h}_i', \mathbf{h}_i'') / \tau)}{\sum_{v\in\mathcal{U}} \exp(s(\mathbf{h}_i', \mathbf{h}_j'') / \tau)},
\end{equation}
where $s(\cdot)$ measures the similarity between two vectors, which is set as cosine similarity function; $\tau$ is the hyper-parameter, known as the temperature in softmax. Analogously, we obtain the contrastive
loss of the item side $\mathcal{L}_{cl}^I$. Combining these two losses, we get the objective function of self-supervised task as $\mathcal{L}_{cl}=\mathcal{L}_{cl}^U+\mathcal{L}_{cl}^I$.


\subsection{Multi-task Training}
\label{section:model-training}
We leverage a multi-task training strategy to optimize the main recommendation task and the auxiliary tasks including translation task, discrimination task and contrastive learning task jointly:
\begin{equation}
  \mathcal{L} = \mathcal{L}_{rec} + \lambda_1 \mathcal{L}_{trans} + \lambda_2 \mathcal{L}_{disc} + \lambda_3 \mathcal{L}_{cl} + \lambda_4 ||\Theta||^2,
\end{equation}
where $\Theta$ is the set of model parameters, $\lambda_1, \lambda_2, \lambda_3, \lambda_4$ are hyperparameters to control the strengths of the diversity preserving loss. $\mathcal{L}_{rec}$ is the loss function of the main recommendation task. In this work, we adopt Bayesian
Personalized Ranking (BPR) loss~\cite{rendle2012bpr}:
\begin{equation}
  \label{eq:l_task_loss}
  \mathcal{L}_{rec} = \sum_{u,i,j\in\mathcal{O}} -\log\sigma(\hat{y}_{u,i} - \hat{y}_{u,j}),
\end{equation}
where $\hat{y}_{u,i}=\mathbf{h}_u^\top \mathbf{h}_i$ is the preference score. $\sigma(\cdot)$ denotes the sigmoid function. $\mathcal{O}=\{(u,i,j)|(u,i)\in\mathcal{O}^+,(u,j)\in\mathcal{O}^-\}$ denotes the training data, and $\mathcal{O}^-$ is the unobserved interactions.

\section{Experiments}
To verify the effectiveness of the proposed {\modelName}, we conduct extensive experiments and report detailed analysis results.

\subsection{Experimental Setups}
\subsubsection{Datasets.}
We evaluate the {\modelName} using three widely-used public benchmark datasets:
Yelp2018\footnote{https://www.yelp.com/dataset}, Amazon-Book\footnote{https://cseweb.ucsd.edu/\textasciitilde jmcauley/datasets.html\#amazon\_reviews} and Movielens-25M\footnote{https://grouplens.org/datasets/movielens/25m/}.
The statistics of these datasets are presented in Table \ref{tab:dataset_info}.

\subsubsection{Evaluation Metrics.}
Due to our focus on Top-N recommendation,
following the convention in the previous research\cite{yu2022self},
we discard ratings less than 4 in Movielens-25M, and reset the rest to 1.
We split the interactions into training, validation, and testing set with a ratio of 7:1:2.
In the test set, we evaluate the performance of each model using the relevancy-based metric Recall@20 and the ranking-aware metric NDCG@20.
\begin{table}[]
  \centering
  \setlength\tabcolsep{5pt} 
  \caption{The statistics of three datasets.}
  \label{tab:dataset_info}
  \begin{tabular}{@{}lllllll@{}}
    \toprule
    Dataset       & \#Users & \#Items & \#Interactions & Density  \\ \midrule
    Yelp2018      & 31,668  & 38,048  & 1,561,406      & 0.1296\% \\
    Amazon-Book   & 52,643  & 91,599  & 2,984,108      & 0.0619\% \\
    Movielens-25M & 155,002 & 27,133  & 3,612,474      & 0.0859\% \\ \bottomrule
  \end{tabular}
\end{table}

\subsubsection{Baselines.}
We compare {\modelName} with other GCN-based recommendation methods, including:
\begin{itemize}
  \item \textbf{LightGCN}\cite{he2020lightgcn} designs a light graph convolution to improve training efficiency and representation ability.
  \item \textbf{SGL}\cite{wu2021self-2} designs an auxiliary tasks via perturbation to the graph structure (such as edge dropout), which achieves greater benifits in long-tail recommendation.
  \item \textbf{NCL}\cite{lin2022improving} improves the recommendation performance by clustering similar nodes to provide semantic neighbors and structural neighbors.
  \item \textbf{RGCF}\cite{tian2022learning} integrates a graph denoising module and a diversity preserving module to enhance the robustness of GCN-based recommendation.
  \item \textbf{SimGCL}\cite{yu2022graph} proves the positive correlation between the uniformity of representations and the ability to debias through feature-level perturbation and contrastive learning, and achieved greater long-tail performance than SGL.
\end{itemize}



\subsubsection{Settings and Hyperparameters.}
We develop the model using the open-source SELFRec\footnote{https://github.com/Coder-Yu/SELFRec} \cite{yu2022self}.
For a fair comparison, we prioritize the hyperparameters reported in the original papers for each baseline when feasible. In cases where this information is not provided, we conduct a grid search to adjust the hyperparameters.
As for the general settings of all the baselines, the Xavier initialization\cite{pmlr-v9-glorot10a} is used on all embeddings.
The embedding size is 64, the parameter for $L_2$ regularization is $10^{-4}$ and the batch size is 2048.
We use Adam\cite{kingma:adam} with the learning rate 0.001 to optimize all the models.
More settings can be found in {https://github.com/im0qianqian/LAGCL}.

\subsection{Performance Comparison}
Table \ref{tab:experiment_result_all} shows our comparison results with other baselines in three datasets.
We have the following observations:
(1) Our proposed {\modelName} consistently outperforms all baselines in different datasets and metrics.
Specifically, {\modelName} achieves a relative improvement of 25.6\%, 61.8\%, and 9.6\%
on Recall@20 compared to LightGCN on the Yelp2018, Amazon Book, and Movielens-25M datasets, respectively.
Compared to the strongest baseline (SimGCL), {\modelName} also achieves better performance, e.g,
about 1.81\%, 2.35\%, 0.73\% performance improvement of Recall@20 on the same datasets.
(2) All graph contrastive learning based methods \eg SGL, NCL, RGCF, SimGCL, show significant improvement compared to LightGCN on three datasets,
which verifies the effectiveness of contrastive learning for collaborative filtering.
SimGCL achieves better performance than other baselines, demonstrating that feature-level augmentation is more suitable for collaborative filtering tasks than structure-level augmentation.
It is worth noting that our method predicts neighborhood information for tail nodes while preserving the original graph structure.
{\modelName} incorporates the advantages of feature-level augmentation and avoids the possibility of drastic changes in tail node information due to graph structural changes caused by structure-level augmentation.
As a result, our method {\modelName} achieves the state-of-the-art performance.


\begin{table}
  \caption{Overall performance comparsion. The percentage in brackets denote the relative performance improvement over LightGCN. The best results are bolded and the best results of baseline are underlined.}
  \setlength\tabcolsep{3pt} 
  \resizebox{\textwidth}{!}{
    \begin{tabular}{l|cc|cc|cc}
      \toprule
      \multirow{2}{*}{Method} & \multicolumn{2}{|c|}{Yelp2018} & \multicolumn{2}{|c|}{Amazon Book} & \multicolumn{2}{|c}{MovieLens-25M}                                                                                         \\
                              & Recall@20                      & NDCG@20                           & Recall@20                          & NDCG@20                     & Recall@20                  & NDCG@20                    \\
      \midrule
      LightGCN                & 0.0583                         & 0.0486                            & 0.0323                             & 0.0254                      & 0.3267                     & 0.2276                     \\
      SGL                     & 0.0659(+13.0\%)                & 0.0541(+11.4\%)                   & 0.0443(+37.0\%)                    & 0.0352(+38.5\%)             & 0.3471(+6.2\%)             & 0.2440(+7.2\%)             \\
      NCL                     & 0.0663(+13.7\%)                & 0.0547(+12.5\%)                   & 0.0426(+32.0\%)                    & 0.0331(+30.2\%)             & 0.3292(+0.8\%)             & 0.2306(+1.3\%)             \\
      RGCF                    & 0.0591(+1.5\%)                 & 0.0487(+0.1\%)                    & 0.0345(+6.9\%)                     & 0.0274(+7.9\%)              & 0.3137(-4.0\%)             & 0.2060(-9.5\%)             \\
      SimGCL                  & \underline{0.0719(+23.4\%)}    & \underline{0.0600(+23.4\%)}       & \underline{0.0510(+57.9\%)}        & \underline{0.0406(+59.8\%)} & \underline{0.3553(+8.8\%)} & \underline{0.2468(+8.4\%)} \\
      \bfseries {\modelName}  & \bfseries 0.0732(+25.6\%)      & \bfseries 0.0604(+24.3\%)         & \bfseries 0.0522(+61.8\%)          & \bfseries 0.0415(+63.4\%)   & \bfseries 0.3579(+9.6\%)   & \bfseries 0.2509(+10.2\%)  \\
      \bottomrule
    \end{tabular}
  }
  \label{tab:experiment_result_all}
\end{table}



\subsection{Ablation study}
\begin{figure}[ht]
  \centering
  \includegraphics[width=0.9\textwidth]{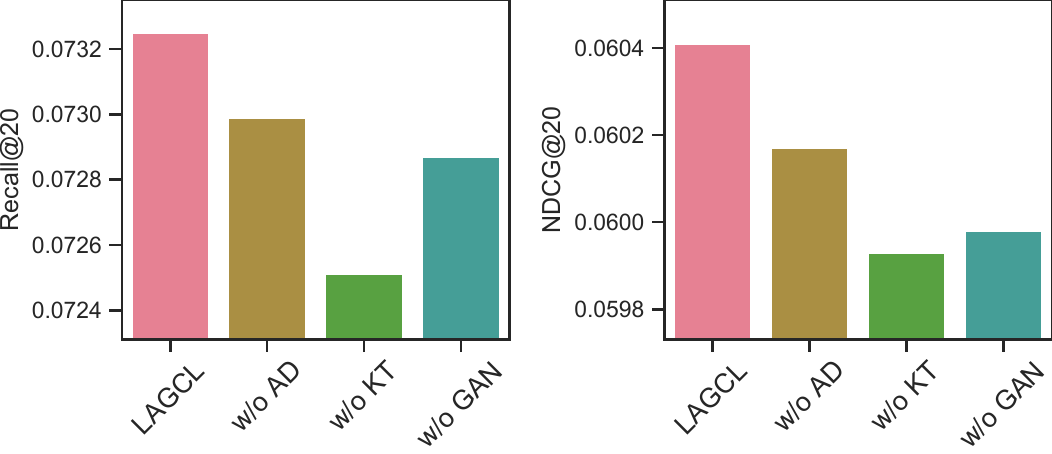}
  \caption{Ablation study in the Yelp2018.}
  \label{fig:ablation_study}
\end{figure}

We conduct an ablation study to compare several ablated variants, including the ``w/o AD" variant that uses random dropout instead of the auto drop module. We also consider the ``w/o KT" variant that do not use knowledge transfer module to augment the tail nodes,
and the ``w/o GAN" variants, which are generative adversarial networks proposed in Section \ref{section:long-tail-aware-contrastive-learning-module}.
We have the following observations from Fig. \ref{fig:ablation_study}. (1) Removing any of the components leads to a performance decrease, with the knowledge transfer module contributing the most. This demonstrates the importance of using knowledge transfer module to augment tail nodes is crucial in GCN-based recommendation scenarios. (2) Using random dropout instead of the auto drop module results in a decrease in performance. It indicates that auto drop module can help to extract informative transition patterns from head nodes that the random dropout strategy cannot learn. (3) Removing the generative adversarial networks significantly decreases the performance. This demonstrates that we cannot achieve meaningful data augmentation without the generative adversarial networks to ensure the distribution match between pseudo-head/tail nodes and real-head/tail nodes.

\subsection{Further Analysis of {\modelName}}
\subsubsection{Distribution uniformity analysis.}

\begin{figure}[ht]
  \centering
  \begin{subfigure}{\textwidth}
    \includegraphics[width=\textwidth]{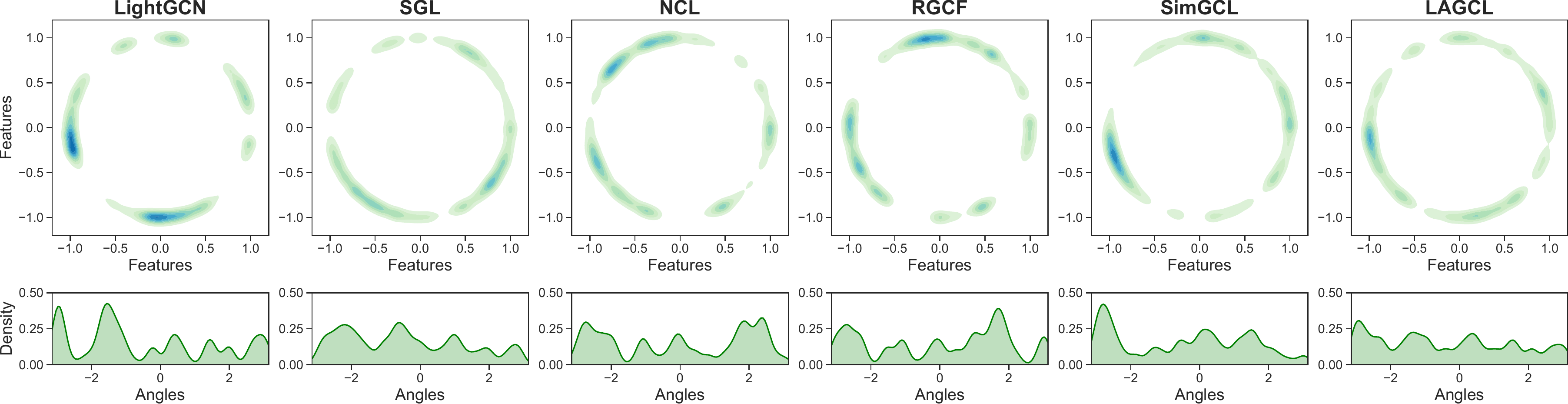}
    \caption{Distribution of user representations learned from Yelp2018 dataset.}
  \end{subfigure}\vspace{1.5mm}
  \begin{subfigure}{\textwidth}
    \includegraphics[width=\textwidth]{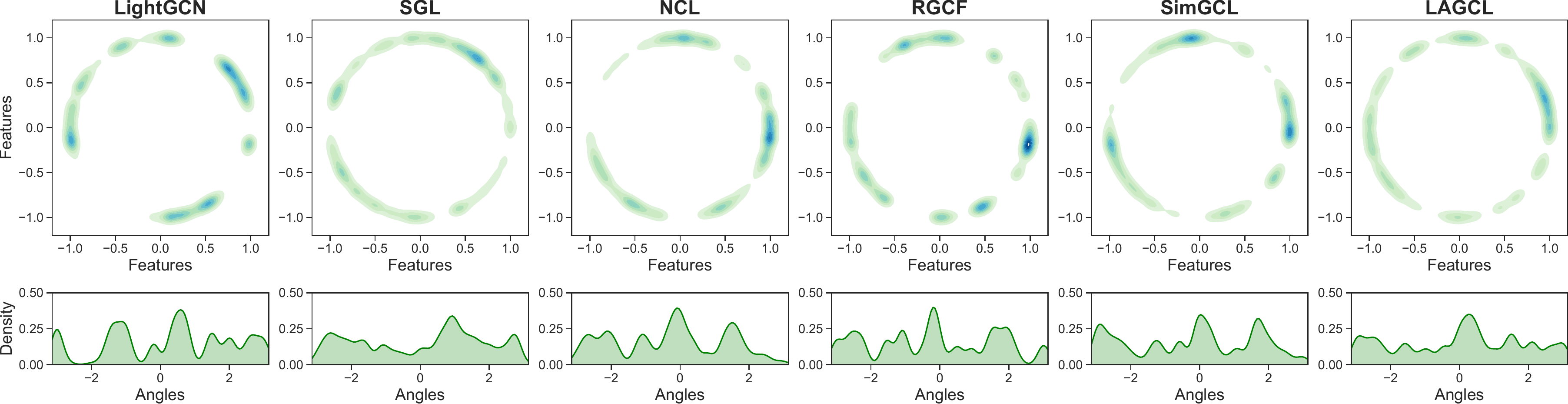}
    \caption{Distribution of item representations learned from Yelp2018 dataset.}
  \end{subfigure}
  \caption{The distribution of user and item representations learned from the Yelp2018 dataset. The top of each figure plots the Gaussian Kernel Density Estimation (KDE) in $\mathbb{R}^2$ (the darker the color is, the more points fall in that area). The bottom of each figure plots the KDE on angles (i.e., $\text{atan2}(y, x)$ for each point $(x,y)\in \mathcal{S}^1$)}
  \label{fig:distribute_cricle}
\end{figure}

A key contribution of the proposed {\modelName} is the utilization of long-tail augmentation to supplement the neighbor information of tail nodes for GCN-based recommendation. To better understand the benefits brought by {\modelName}, we visualize the learned embeddings in Fig~\ref{fig:distribute_cricle} to illustrate how the proposed approach affects representation learning. 
We use Gaussian kernel density estimation (KDE)~\cite{10.1214/10-AOS799} to plot user and item embedding distributions in two-dimensional space. Additionally, we visualize each node's density estimations on angles on the unit hypersphere $\mathcal{S}^1$ (i.e., circle with radius 1).
We can see that, the embeddings learned by LightGCN fall into serveral clusters located on narrow arcs.
Graph contrastive learning-based methods exhibit a more uniform distribution than LightGCN, where SimGCL has a more uniform distribution than other structure-based graph contrastive learning-based methods (SGL, NCL and RGCF).
When compared to the best baseline SimGCL, the {\modelName} distribution has fewer dark areas on the circle, indicating that {\modelName} has a more uniform distribution that benefits the model performance. Previous studies~\cite{wang2020understanding,yu2022graph} have shown a strong correlation between contrastive learning and the uniformity of learned representations. Thus, we speculate that a more uniform distribution of embeddings could endow the model with better capacity to capture diverse user preferences and item characteristics.


\subsubsection{Long Tail Degree Analysis.}
\begin{figure}[ht]
  \centering
  \begin{subfigure}{.4\textwidth}
    \includegraphics[width=\textwidth]{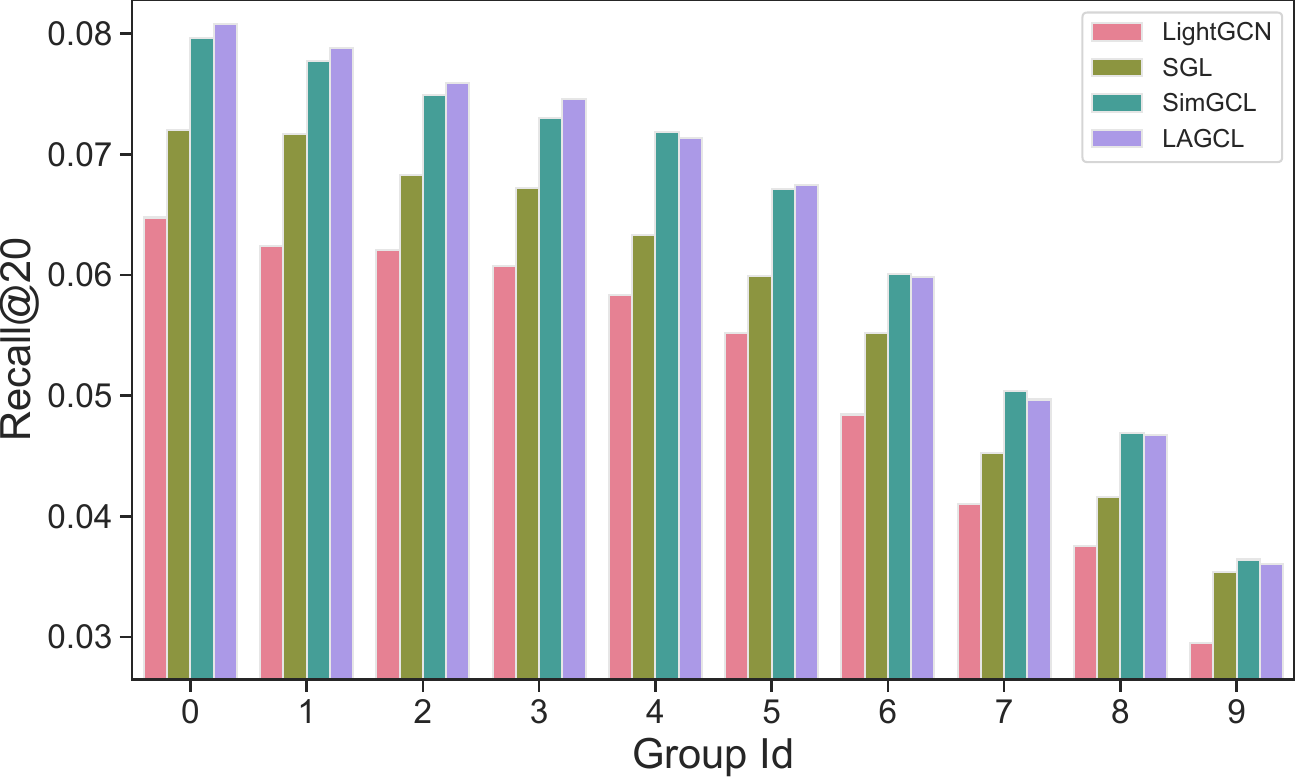}
    \caption{Yelp2018}
  \end{subfigure}
  \begin{subfigure}{.4\textwidth}
    \includegraphics[width=\textwidth]{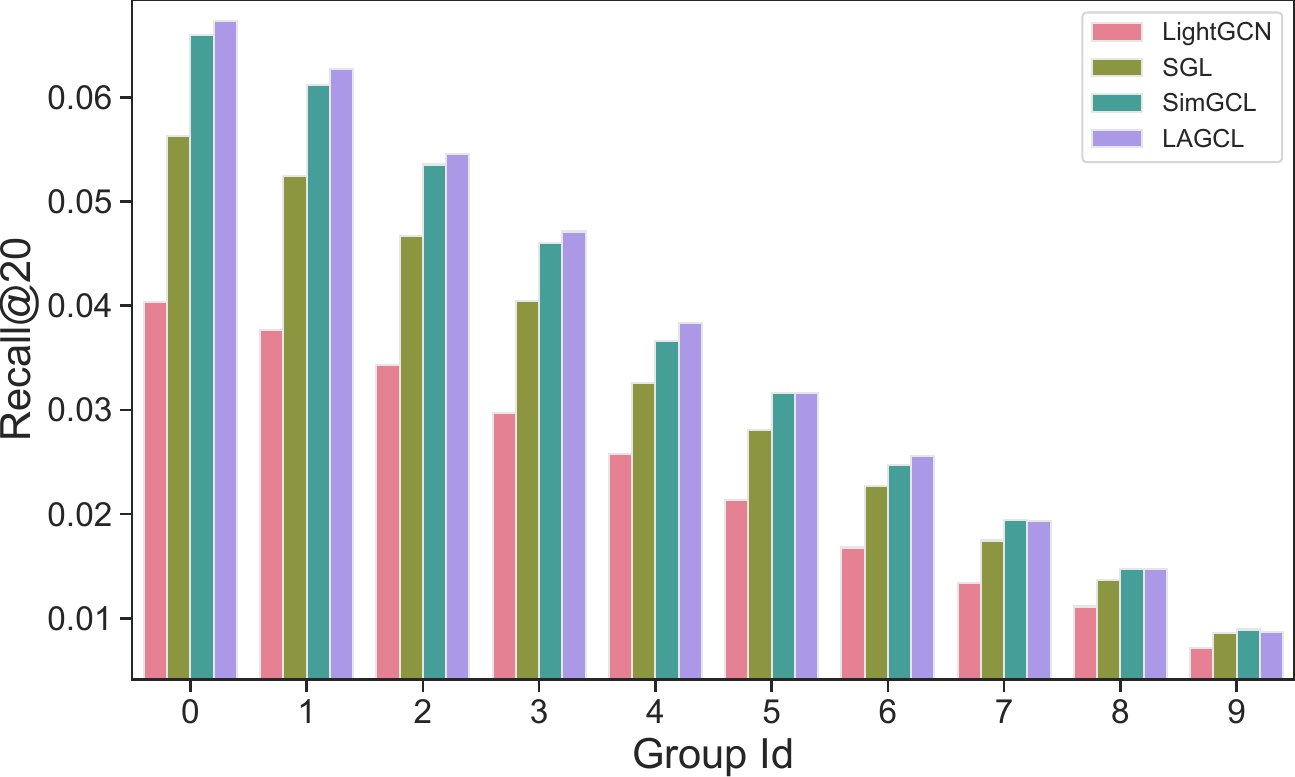}
    \caption{Amazon Book}
  \end{subfigure}
  \caption{Performance comparison of different user groups. In addition to the classic method LightGCN, we also select two different approaches to alleviate the long-tail distribution in the graph, namely SGL and SimGCL.}
  \label{fig:diff_user_group_performance}
\end{figure}

To verify whether {\modelName} can provide additional performance gains for tail nodes,
we divide each user into 10 groups of equally size based on their node degree in the user-item bipartite graph,
as shown in Fig. \ref{fig:diff_user_group_performance}.
The smaller the Group Id, the lower node degree, and the lower user activity.
We compare {\modelName} with other baselines that alleviate long-tail distribution in graph,
such as SGL employs structure-level augmentation and SimGCL utilizes feature-level augmentation.
The results show that the performance of these methods are very similar for high-activity users,
while for low-activity users, {\modelName} exhibits better performance.
This proves that our method has significant gains for modeling tail users.

\subsubsection{Degree threshold $k$.}
\begin{figure}[ht]
  \centering
  \begin{subfigure}{.4\textwidth}
  \includegraphics[width=\textwidth]{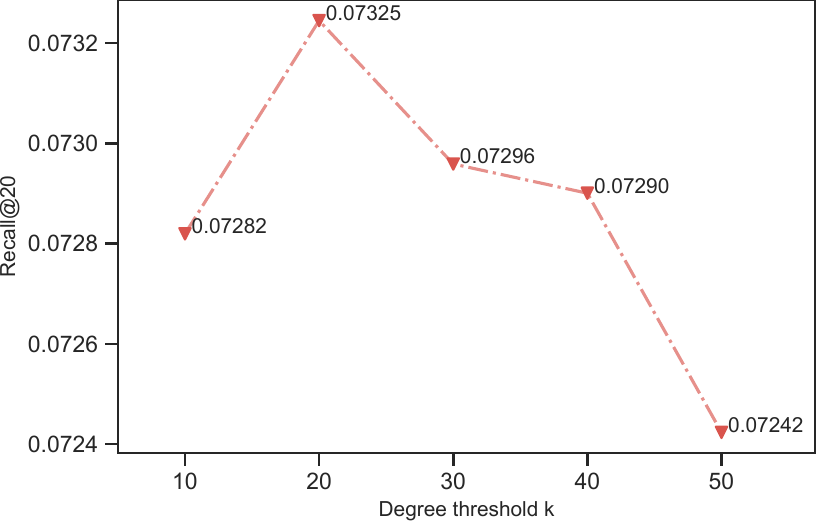}
  \end{subfigure}
  \begin{subfigure}{.4\textwidth}
    \includegraphics[width=\textwidth]{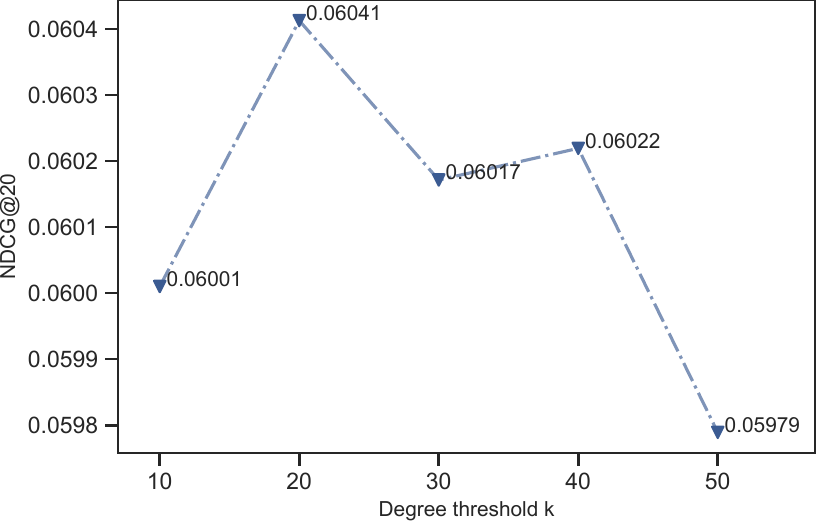}
  \end{subfigure}
  \caption{Performance of {\modelName} in the Yelp2018 when adjusting the degree threshold k.}
  \label{fig:parameter_sensitivity_analysis}
\end{figure}

As shown in Fig. \ref{fig:parameter_sensitivity_analysis},
we conduct a parameter sensitivity experiment on the threshold for dividing the head and tail nodes using the Yelp2018 dataset.
The results show that our model achieve the best performance at $k=20$.
We conclude that:
(1) {\modelName} can learn a transfer strategy through the head users to help bring benefits to tail users.
(2) A larger $k$ will result in fewer number of the head users, which may cause the knowledge transfer module to be insufficiently learned.
Conversely, a smaller $k$ will result in significant quality differences within the head users, causing inaccurate learning in the knowledge transfer module.
Therefore, choosing a suitable threshold for division can maximize the overall benefit.






\section{Related Work}
\subsection{GCN-based Recommendation}
GCN-based recommendation methods have treated interaction data as a bipartite graph and applied graph neural networks to obtain embeddings while capturing high-order information. For example, GC-MC~\cite{berg2017graph} applies the Graph Convolution Network (GCN) on the user-item graph and employs one convolutional layer to exploit the direct connections between users and items. PinSage~\cite{ying2018graph} is an industrial solution that leverages efficient random walks and graph convolutions to generate item embeddings in Pinterest. NGCF~\cite{wang2019neural} models high-order connectivities by effectively injecting the collaborative signal into the embedding process explicitly. LightGCN~\cite{he2020lightgcn} simplifies the design of GCN by removing feature transformation and nonlinear activation, making it more effective for recommendation purpose. 

However, all of these works perform training in the supervised learning paradigm, which heavily relies on labeled data. Since the observed interactions are considerably sparse compared to the entire interaction space, they may not be sufficient to acquire effective representations.

\subsection{Self-supervised Learning in Recommender Systems}
Self-supervised learning is an emerging technique that leverages unlabeled data to achieve impressive success in various fields, including computer vision~\cite{chen2020simple,he2020momentum}, natural language processing~\cite{devlin2018bert,lan2019albert} and graph learning~\cite{wu2021self,liu2022graph,xie2022self}. Inspired by the success of these works, self-supervised learning has also been applied to recommender systems, where it has shown great improvement~\cite{yu2022self}. For instance, $S^3$-Rec utilizes four self-supervised objectives to learn the correlations among attribute, item, subsequence, and sequence by maximizing the mutual information~\cite{zhou2020s3}. Similarly, CLCRec further utilizes contrastive learning to mitigate the cold-start issue by maximizing the mutual dependencies between item content and collaborative signals~\cite{wei2021contrastive}. Moreover, a promising line of research has incorporated contrastive learning into graph-based recommenders to tackle the label sparsity issue with self-supervision signals. For example, SGL~\cite{wu2021self-2} uses node dropout, edge dropout and random walk on the user-item interaction graph to generate different views and maximize the agreement between them. NCL~\cite{lin2022improving} incorporates structural and semantic neighbors to enhance graph-based collaborative filtering. SimGCL~\cite{yu2022graph} generates contrastive views by adding uniform noise in the embedding space instread of graph augmentations.


\section{Conclusion}

In this work, we propose a novel long-tail augmented graph contrastive learning (LAGCL) method for recommendation. 
Specifically, we introduce a learnable long-tail augmentation schema that enhances tail nodes by supplementing them with predicted neighbor information. To make the data augmentation schema learnable, we design an auto-drop strategy to generate pseudo-tail nodes from head nodes and a knowledge translation module to reconstruct the head nodes from pseudo-tail nodes. To ensure the effectiveness of the data augmentation, we leverage adversarial constraints to distinguish between pseudo-tail and real-tail nodes, as well as between augmented tail nodes and real-head nodes. Comprehensive experiments demonstrate the effectiveness of our proposed {\modelName}.

\section*{Ethics Statement}
Our work aims to improve the accuracy of recommendations using only exposure and click data, without collecting or processing any personal information. We recognize the potential ethical implications of recommendation systems and their impact on user privacy, and we have taken steps to ensure that our work adheres to ethical standards.

We understand that recommendation systems have the potential to influence user behavior and may raise concerns related to user privacy. To address these concerns, we have designed our approach to rely solely on non-personal data, ensuring that our methods do not infringe on user privacy or rights.

In summary, our work focuses on improving recommendation accuracy while upholding ethical standards related to user privacy. We believe that our approach can contribute to the development of recommendation systems that are both effective and ethical.
%
%
%
%
\bibliographystyle{splncs04}
\bibliography{ref}

\newpage
\appendix
\end{document}